\documentclass[submission, LectureNotes, a4paper]{SciPost}

\usepackage{hyperref}
\usepackage[english]{babel}
\usepackage{amsmath,amssymb,slashed,bbm,bbold,siunitx}
\usepackage{ctable,multirow}
\usepackage{graphicx,color}
\usepackage[skipabove=2mm,skipbelow=2mm]{mdframed}
\usepackage[absolute]{textpos}
\usepackage{makeidx}
\usepackage[comma,square,numbers,sort&compress]{natbib}
\usepackage{feynmp}
\usepackage{cleveref}

\newcommand{\bra}[1]{\ensuremath{\langle #1 |}}     
\newcommand{\ket}[1]{\ensuremath{| #1 \rangle}}     
\newcommand{\sprod}[2]{\ensuremath{\left\langle #1 |%
                     #2 \right\rangle}}             

\newcommand{\diag}{{\rm diag}}             

\newcommand{\im}{\textrm{Im}}
\newcommand{\re}{\textrm{Re}}

\begin{document}

\begin{center}{\Large \textbf{
Sterile Neutrinos as Dark Matter Candidates
}}\end{center}

\begin{center}
J.\ Kopp\textsuperscript{*}
\end{center}

\begin{center}
Theoretical Physics Department, CERN, Geneva, Switzerland and \\
Johannes Gutenberg University Mainz, 55099 Mainz, Germany
\\
* jkopp@cern.ch
\end{center}

\begin{center}
\today
\end{center}


\section*{Abstract}
{\bf
In these brief lecture notes, we introduce sterile neutrinos as dark matter candidates.
We discuss in particular their production via oscillations, their radiative decay,
as well as possible observational signatures and constraints.
}

\vspace{10pt}
\noindent\rule{\textwidth}{1pt}
\tableofcontents\thispagestyle{fancy}
\noindent\rule{\textwidth}{1pt}
\vspace{10pt}

\section{Neutrino Masses and Mixings}

In the simplest extension of the Standard Model (SM) that admits neutrino masses,
neutrinos have the following interaction and mass terms:
\begin{align}
  \mathcal{L} &= \sum_{\alpha=e,\mu,\tau} \bigg[
      \frac{g}{\sqrt{2}} \Big( \overline{\nu_{\alpha,L}} \gamma^\rho e_{\alpha,L}
                                                      W_\rho^+ + h.c. \Big)
    + \frac{g}{2\cos\theta_w} \overline{\nu_{\alpha,L}} \gamma^\rho \nu_{\alpha,L} Z_\rho \bigg]
          \nonumber\\
   &- \sum_{\alpha,\beta=e,\mu,\tau} \Big(
        m_{\alpha\beta} \, \overline{\nu_{\alpha,L}} \nu_{\beta,R}  +  h.c. \Big) \,.
  \label{eq:L-weak}
\end{align}
Here, $g$ is the weak coupling constant and $\theta_w$ is the Weinberg angle.
Note that only left-handed neutrinos couple to the weak gauge bosons $W^\pm$ and $Z$.
In terms of Feynman diagrams, the neutrino interaction vertices can be written as
\begin{center}
  \includegraphics[width=12cm]{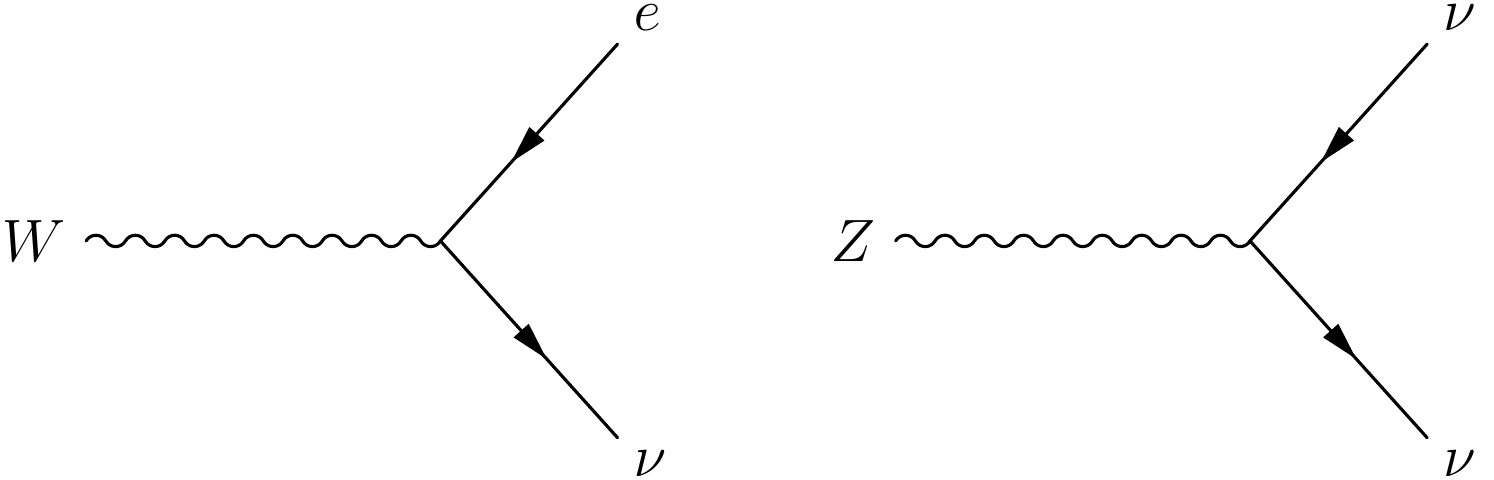}
\end{center}
Note that the mass term $\sum_{\alpha,\beta=e,\mu,\tau} m_{\alpha\beta}
\, \overline{\nu_{\alpha,L}} \nu_{\beta,R}$ in eq.~\eqref{eq:L-weak} is in general
\emph{off-diagonal} (i.e.\ $m_{\alpha\beta}$ can be non-zero even if $\alpha
\neq \beta$). This means that the \emph{flavor eigenstates} or \emph{interaction
eigenstates} $\nu_\alpha$ ($\alpha = e,\mu,\tau$) do not have
a definite mass.

Adding a sterile neutrino $\nu_s$ -- i.e.\ a SM singlet fermion -- to this Lagrangian is
straightforward. $\nu_s$ appears in the neutrino mass term, but not in the weak interaction
term.  The mass term then changes into
\begin{align}
  \mathcal{L}_m = -\sum_{\alpha,\beta=e,\mu,\tau,s} \Big(
    m_{\alpha\beta} \, \overline{\nu_{\alpha,L}} \nu_{\beta,R}  +  h.c. \Big) \,.
\end{align}

The mass matrix $m$ can be diagonalized according to
\begin{align}
  m = U m_D V^\dag \,,
\end{align}
where $m_D = \diag(m_1, m_2, m_3, m_4)$ is a diagonal matrix and $U$, $V$
are unitary matrices.  We define the fields in the neutrino \emph{mass eigenstate}
basis according to
\begin{align}
  \nu_{j,L} &\equiv \sum_\alpha U_{\alpha j}^* \nu_{\alpha,L} \\
  \nu_{j,R} &\equiv \sum_\alpha V_{\alpha j}^* \nu_{\alpha,R} \,.
\end{align}
In terms of the mass eigenstates, the Lagrangian \eqref{eq:L-weak} can be
written as
\begin{align}
  \mathcal{L} &= \sum_{\alpha=e,\mu,\tau} \sum_{j=1,2,3,4}
      \frac{g}{\sqrt{2}} \Big( \overline{\nu_{j,L}} U_{\alpha j}^* \gamma^\rho e_{\alpha,L}
                                                        W_\rho^+ + h.c. \Big)
          \nonumber\\
    &+ \sum_{\alpha=e,\mu,\tau} \sum_{j,k=1,2,3,4}
        \frac{g}{2\cos\theta_w} \overline{\nu_{j,L}} U_{\alpha j}^* \gamma^\rho
                                                     U_{\alpha k} \nu_{k,L} Z_\rho
          \nonumber\\
    &- \sum_{j=1,2,3,4} \Big( m_j \, \overline{\nu_{j,L}} \nu_{j,R}  +  h.c. \Big) \,.
  \label{eq:L-weak-mass}
\end{align}
Thus, a charged current neutrino interaction produces a superposition of
mass eigenstates, for instance
\begin{center}
  \includegraphics[width=12cm]{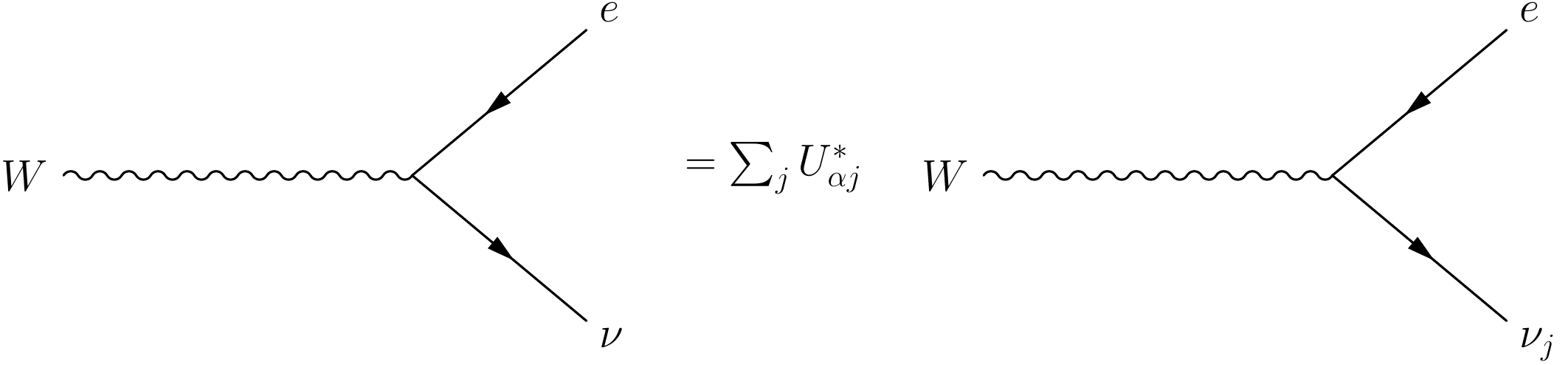}
\end{center}

\section{Neutrino Oscillations}

Since neutrino flavor eigenstates---the states that are produced by the weak
interaction---are superpositions of mass eigenstates---the states with well-defined
kinematics and propagators---we expect quantum interference effects in neutrino
experiments. These effects are the neutrino oscillations.

We start from the weak interaction Lagrangian in the mass basis, \cref{eq:L-weak-mass}.
It implies that the neutrino state of flavor $\alpha$ ($\alpha = e$, $\mu$, $\tau$, $s$)
produced in a weak interaction can be written as the following superposition
of mass eigenstates
\begin{align}
  \ket{\nu_\alpha} = \sum_j U_{\alpha j}^* \ket{\nu_j} \,.
  \label{eq:osc-nu-produced}
\end{align}
Note that even though the transformation of the field operators is
$\nu_{\alpha,L} = \sum_j U_{\alpha j} \nu_{j,L}$, the transformation of the ket-states
is determined by $U^\dag$ rather than $U$. The reason is that these states
are produced by the creation operator $\sum_j U_{\alpha j}^* \bar{\nu}_{j,L}$,
not by the annihilation opearators $\sum_j U_{\alpha j} \nu_{j,L}$.
Treating $\ket{\nu_j}$ as plane wave states, the wave function at a distance
$L$ from the production point, and at a time $T$ after production, is given
by
\begin{align}
  \ket{\nu_\alpha(T, L)} = \sum_j U_{\alpha j}^* e^{-i E_j T + i p_j L} \ket{\nu_j} \,.
  \label{eq:osc-nu-evolved}
\end{align}
Note that the energy $E_j$ and the momentum $p_j$ are in general different for
the different mass eigenstates because the kinematics of the production process
is different for different mass.

A neutrino detector measures the neutrino flavor, i.e.\ it detects the
neutrino in a state
\begin{align}
  \bra{\nu_\beta} = \sum_j U_{\beta j} \bra{\nu_j} \,.
  \label{eq:osc-nu-detected}
\end{align}
Therefore, the amplitude for a neutrino produced as $\ket{\nu_\alpha}$ to be
detected as $\bra{\nu_\beta}$ is
\begin{align}
  \sprod{\nu_\beta}{\nu_\alpha(T, L)}
    &= \sum_{j,k} U_{\alpha j}^* U_{\beta k} e^{-i E_j T + i p_j L} \sprod{\nu_k}{\nu_j} \\
    &= \sum_j U_{\alpha j}^* U_{\beta j} e^{-i E_j T + i p_j L} \,.
  \label{eq:osc-A-1}
\end{align}
The \emph{oscillation probability} is thus
\begin{align}
  P_{\alpha\beta}(T,L) = |\sprod{\nu_\beta}{\nu_\alpha(T, L)}|^2
                       = \sum_{j,k} U_{\alpha j}^* U_{\beta j} U_{\alpha k} U_{\beta k}^*
                           e^{-i (E_j -E_k) T + i (p_j - p_k) L} \,.
  \label{eq:osc-P-1}
\end{align}
In a typical neutrino oscillation experiment, we do not know when precisely each
neutrino is produced (the experimental uncertainty in the production time is much
larger than the energy uncertainty of each individual neutrino). Therefore, we
should integrate over $T$:\footnote{This approach may not be strictly valid
any more for long-baseline oscillation experiments which typically have excellent
timing resolution due to short beam spills and good detector resolution. A more
rigorous calculation describing the neutrino as a wave packet confirms, however,
that the expression for the oscillation probability remains correct even for
long-baseline oscillation experiments.}
\begin{align}
  P_{\alpha\beta}(L)
    &= \frac{1}{N} \int \! dT \, P_{\alpha\beta}(T,L) \\
    &= \frac{1}{N} \sum_{j,k} U_{\alpha j}^* U_{\beta j} U_{\alpha k} U_{\beta k}^*
         \exp \Big[ i \Big(\sqrt{E^2 - m_j^2} - \sqrt{E^2 - m_k^2}\Big) L \Big] \,
         2\pi \delta(E_j - E_k) \\
    &\simeq \sum_{j,k} U_{\alpha j}^* U_{\beta j} U_{\alpha k} U_{\beta k}^*
         \exp \bigg[ -i \frac{\Delta m_{jk}^2 L}{2 E} \bigg] \,.
    \label{eq:osc-P-2}
\end{align}
Here, $N$ is a normalization constant, which is chosen such that
$\sum_\beta P_{\alpha\beta}(L) = 1$.
In the last line of eq.~\eqref{eq:osc-P-2}, we have made the approximation
$|m_j^2 - m_k^2| \ll E^2 = E_j^2 = E_k^2$ (equal energy approximation)
and carried out a Taylor expansion in the mass squared difference
\begin{align}
  \Delta m_{jk}^2 \equiv m_j^2 - m_k^2 \,.
  \label{eq:dmsq}
\end{align}
We could also have made the (somewhat
unjustified) assumption that all neutrino mass eigenstates are emitted with
the same momentum $p$, but different energies. This assumption would have led to
the same result, but with phase factor $\exp[-i \Delta m_{jk}^2 T / (2E)]$
instead of $\exp[-i \Delta m_{jk}^2 L / (2E)]$. Since neutrinos travel
at the speed of light (up to negligible corrections of order $\Delta m_{jk}^2
/ E^2$), we can set $L = T$, so that the two approaches become completely
equivalent.

The expression for $P_{\alpha\beta}(L)$ becomes particularly simple in the
2-flavor approximation, where the mixing matrix $U$ can be written as
\begin{align}
  U = \begin{pmatrix}
        \cos\theta & \sin\theta \\
       -\sin\theta & \cos\theta
      \end{pmatrix} \,.
\end{align}
For instance, if the two flavors are $e$ and $\mu$, we obtain
\begin{align}
  P_{e\mu}^\text{2-flavor}(L)
    &= |U_{e1}|^2 |U_{\mu 1}|^2 + |U_{e2}|^2 |U_{\mu 2}|^2 \nonumber\\
    &\hspace{1cm} + U_{e 1} U_{\mu 1} U_{e 2} U_{\mu 2} \bigg[
         \exp\bigg( -i \frac{\Delta m^2 L}{2 E} \bigg)
       + \exp\bigg( +i \frac{\Delta m^2 L}{2 E} \bigg)
       \bigg] \\
    &= 2 \cos^2\theta \sin^2\theta
         - 2\cos^2\theta \sin^2\theta \cos\bigg[ \frac{\Delta m^2 L}{2 E} \bigg] \\
    &= \frac{1}{2} \sin^2 2\theta
          \bigg( 1 - \cos\bigg[ \frac{\Delta m^2 L}{2 E}\bigg] \bigg) \\
    &= \sin^2 2\theta \, \sin^2 \bigg[ \frac{\Delta m^2 L}{4 E} \bigg] \,.
\end{align}

Several comments are in order here:
\begin{itemize}
  \item States with different energy and momentum $(E_j, p_j)$, $j=1,2,3$
    can interfere only if the energy and momentum uncertainties associated with the
    production and detection processes are larger than $|E_j - E_k|$, $|p_j -
    p_k|$.  This is always satisfied in practice as the typical momentum
    uncertainty associated with a neutrino production process is at least of
    the order of an inverse interatomic distance, i.e.\ of order keV.
    Therefore, the interference conditions for different neutrino mass
    eigenstates is easily satisfied.

  \item A related point: since the energy and momentum uncertainties are so
    important for interference to happen, it is not correct to treat neutrinos
    as plane waves. A wave packet formalism is more appropriate.

  \item The approximation $\sqrt{E^2 - m_j^2} - \sqrt{E^2 - m_k^2} \simeq
    -\Delta m_{jk}^2 / (2 E)$ does not require $m_j, m_k \ll E$, but only
    $m_j^2 - m_k^2 \ll E^2$.  (This is sufficient for writing
    $m_j^2 = m_k^2 + \Delta m_{jk}^2$ and then expanding in $\Delta m_{jk}^2$.)

  \item For antineutrinos, the above derivation goes through in exactly
    the same way, except that $U$ should be replaced by $U^*$ everywhere.
    This is because an antineutrino is created by the field operator
    $\nu$ rather than the operator $\bar\nu$, and the corresponding weak
    interaction term in the Lagrangian \eqref{eq:L-weak} is the hermitian
    conjugate of the term creating neutrinos.  We denote oscillation
    probabilities for $\bar\nu_\alpha \to \bar\nu_\beta$ transitions
    by $P_{\bar\alpha \bar\beta}(L)$.
\end{itemize}

\section{Sterile Neutrinos as Dark Matter Candidates}

Sterile neutrinos with masses $> \text{keV}$ have all the properties required
to account for the DM in the Universe: they are electrically neutral, become
non-relativistic early on (thus forming cold dark matter), can have very weak
couplings with other particles (if the relevant mixing angles are small), and
are stable over cosmological time scales.

An important question for any DM candidate is how the DM abundance observed in
the Universe is determined. For the case of sterile neutrinos, the minimal
mechanism is the \emph{Dodelson--Widrow} mechanism~\cite{Dodelson:1993je},
which we will outline now.

The assumption is that, very early on, no sterile neutrinos exist. Later, they
are produced via active-to-sterile ($\nu_a \to \nu_s$) neutrino oscillations.
For $\mathcal{O}(\text{keV})$ masses, the oscillation length/oscillation time
scale $L^\text{osc} = 4\pi E / \Delta m^2$ is very small,
so a $\nu_a$--$\nu_s$ superposition is produced very
quickly. In the subsequent discussion, is is therefore sufficient to
consider the averaged effect of oscillations.
For small mixing angle, the neutrino state then consists mostly of $\nu_a$, with a
small ($\sim \frac{1}{2} \sin^2 2\theta$)
admixture of $\nu_s$. Neutrino collisions
with other particles act as quantum mechanical ``measurements'', collapsing the
wave function either into $\nu_s$ (with a probability of $\frac{1}{2} \sin^2
2\theta$), or into $\nu_a$ (with a probability of $1 - \frac{1}{2} \sin^2
2\theta$).  Afterwards, oscillations start again. Active neutrinos again
acquire a $\nu_s$ component $\sim \frac{1}{2} \sin^2 2\theta$, and sterile
neutrinos acquire a $\nu_a$ component of the same magnitude. However, since
$\nu_s$ are much less abundant than $\nu_a$, the back-conversion is negligible.
After many collisions, the sterile neutrino abundance has increased to the
level observed today. Eventually, collisions cease because the primordial gas
becomes too diluted, and the $\nu_s$ abundance present at this time ``freezes
in''.  Note that, before freeze-in, active neutrinos are continuously replenished
via pair production or charged current interactions.

Dodelson--Widrow production of sterile neutrinos is described by the Boltzmann
equation
\begin{align}
  \bigg( \frac{\partial}{\partial t}
    - H \, E \, \frac{\partial}{\partial E} \bigg) f_s(E, t)
  = \bigg[ \frac{1}{2} \sin^2 (2\theta_M(E,t)) \Gamma(E,t) \bigg] f_a(E, t) \,,
  \label{eq:Boltzmann-Dodelson-Widrow}
\end{align}
where $f_s(E, t)$ and $f_a(E, t)$ are the time-dependent momentum distribution
functions of sterile and active neutrinos, respectively, and $H$ is the Hubble
parameter. Before the
interactions between active neutrinos and other SM particles freeze out
(the epoch relevant here because the mechanism relies on these collisions),
$f_a(E, t)$ is just a Fermi--Dirac distribution
\begin{align}
  f_a(E, t) = \frac{1}{e^{E/T} + 1} \,.
\end{align}
The quantity
\begin{align}
  \Gamma(E, t) \simeq \frac{7\pi}{24} G_F^2 T^4 E
\end{align}
in eq.~\eqref{eq:Boltzmann-Dodelson-Widrow} is the active neutrino interaction
rate.  The expression in square brackets is thus the probability for the
neutrino state to collapse to $\nu_s$ in a collision.\footnote{It may seem odd
that the neutrino can collapse into $\nu_s$ even though only $\nu_a$
interact.  This paradox can only be resolved in a more careful treatment of the
Dodelson--Widrow mechanism using the density matrix formalism.} $\theta_M(E,
t)$ denotes the mixing angle in matter.  The second term on the left hand side
of eq.~\eqref{eq:Boltzmann-Dodelson-Widrow} describes the change in the energy
spectrum due to redshift. Indeed, we have
\begin{align}
  \frac{d}{dt} f_s(E,t) = \bigg( \frac{\partial}{\partial t}
    + \frac{dE}{dt} \frac{\partial}{\partial E} \bigg) f_s(E, t) \,,
\end{align}
and, for relativistic neutrinos,
$dE/dt = d(E_0 a^{-1}) / dt = -E_0 a^{-2} \dot{a} = -H\,E$, with $a$ the
scale factor of the Universe and $H = \dot{a} / a$. It is justified to
assume relativistic neutrinos here as Dodelson--Widrow production peaks
at a temperature of order \SI{100}{MeV}.

From eq.~\eqref{eq:Boltzmann-Dodelson-Widrow}, we can compute an evolution
equation also for the ratio of number densities of sterile and active
neutrinos, $r(t) \equiv n_s(t) / n_a(t)$, with $n_i(t) = 2 \int d^3p \, f_i(E,t)
/ (2\pi)^3$.  In doing so, it is convenient to go from derivatives with respect
to $t$ to derivatives with respect to $a(t)$. We use
\begin{align}
  \frac{d}{da} n_s(t)
  &= \frac{d}{da} 2 \int\!\frac{d^3p}{(2\pi)^3} f_s(E, t) \\
  &= 2 \frac{d}{da} \int\!\frac{4\pi E^2 dE}{(2\pi)^3} f_s(E, t) \\
  &= \frac{2}{\dot{a}} \int\!\frac{4\pi E^2 dE}{(2\pi)^3} \frac{\partial}{\partial t} f_s(E, t)
    - 2 \int\!\frac{4\pi E^2 dE}{(2\pi)^3} \frac{E}{a} \frac{\partial}{\partial E} f_s(E, t)
    - 6 \int\! \frac{4\pi E\,dE}{(2\pi)^3} \frac{E}{a} f_s(E, t) \,,
\end{align}
or, equivalently,
\begin{align}
  \dot{a} \frac{d}{da} n_s
  &= 2 \int\!\frac{4\pi E^2\,dE}{(2\pi)^3} \frac{\partial}{\partial t} f_s(E, t)
    - 2 \int\!\frac{4\pi E^2 dE}{(2\pi)^3} H \, E \frac{\partial}{\partial E} f_s(E, t)
    - 3 H n_s \,.
\end{align}
We can thus rewrite eq.~\eqref{eq:Boltzmann-Dodelson-Widrow} as
\begin{align}
  \dot{a} \frac{d}{da} n_s + 3 H n_s  =  \gamma  n_a \,,
\end{align}
where we have defined
\begin{align}
  \gamma \equiv \frac{1}{n_a} \int\! \frac{d^3p}{(2\pi)^3}\,
    \sin^2 (2\theta_M) \Gamma(E,t) \frac{1}{e^{p/T} + 1} \,.
\end{align}
Since, moreover,
\begin{align}
  \frac{d}{da} n_a = -\frac{3}{a} n_a \,,
\end{align}
we obtain
\begin{align}
  \dot{a} \frac{d}{da} r + \dot{a} \frac{r}{n_a} \frac{d}{da} n_a
    + 3 H r(t) &=  \gamma \,, \\
  \Leftrightarrow\quad
  \dot{a} \frac{d}{da} r  &=  \gamma \,, \\
  \Leftrightarrow\quad
  a H \frac{d}{da} r  &= \gamma \,, \\
  \Leftrightarrow\quad
  \frac{dr}{d\ln a} &= \frac{\gamma}{H} \,.
  \label{eq:Boltzmann-DW-r}
\end{align}
Note that, in the above derivation, we have neglected the time-dependence of
the effective number of relativistic degrees of freedom, $g_*$.
At epochs where $g_*$ changes, the dependence of energy on the scale factor is no longer
simply $E \propto a^{-1}$ because the energy of degrees of freedom that
disappear is distributed among those remaining in thermal equilibrium.
When this is taken into account, eq.~\eqref{eq:Boltzmann-Dodelson-Widrow}
turns into~\cite{Dodelson:1993je}
\begin{align}
  \frac{d}{d\ln a} r &= \frac{\gamma}{H} + r \frac{d}{d\ln a} g_*\,. 
\end{align}

\section{Sterile Neutrino Decay}

We have mentioned above that any DM candidate should be stable over cosmological
timescales. For sterile neutrinos this is the case in the sense that
they live much longer than the age of the Universe if their mixing angles with
active neutrinos are sufficiently small. But they are not absolutely stable.
A massive, mostly sterile neutrino $\nu_4$ with a small admixture of a light,
mostly active neutrino state $\nu_1$ can decay through the following
diagrams:
\begin{center}
  \includegraphics[width=\textwidth]{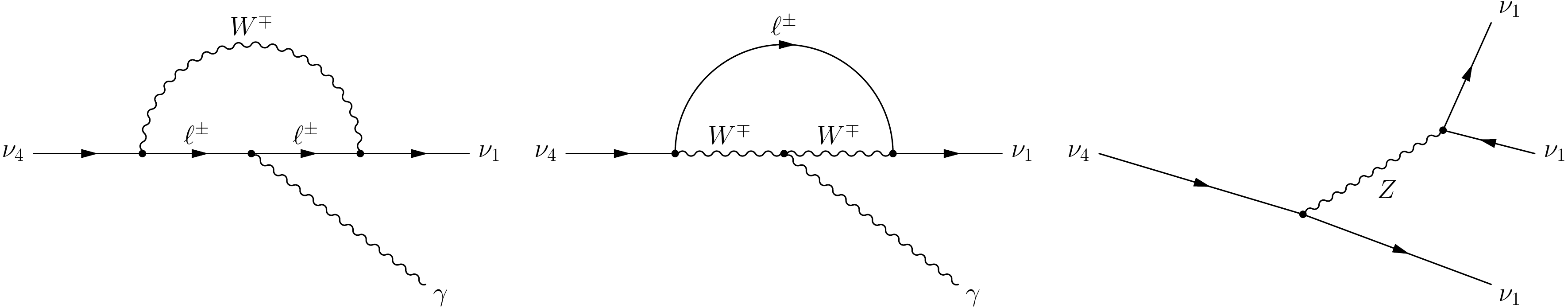}
\end{center}
The third of these is phenomenologically irrelevant because the decay products
are invisible.  It can only be used to impose the constraint that the lifetime
of $\nu_4$ should be much larger than the age of the Universe to provide a
successful DM candidate.  The first two diagrams, on the other hand,
lead to radiative neutrino decay $\nu_4 \to \nu_1 \gamma$.

The rate for for $\nu_4 \to \nu_a + \gamma$ is \cite{Shrock:1974nd,
Lee:1977tib, Marciano:1977wx, Pal:1981rm, Shrock:1982sc}
\begin{align}
  \Gamma(\nu_4 \to \nu_a \gamma)^\text{D}
  &= \frac{9 \alpha_\text{em} G_F^2 m_4^5}{512 \pi^4}
       \sum_{j=1,2,3} \bigg( 1 - \frac{m_j^2}{m_4^2} \bigg)^3
       \bigg( 1 + \frac{m_j^2}{m_4^2} \bigg)
       \bigg| \sum_{\alpha=e,\mu,\tau}
         \bigg( 1 - \frac{m_\alpha^2}{2 M_W^2} \bigg) U_{\alpha 4} U_{\alpha j}^* \bigg|^2
                                   \nonumber\\[0.2cm]
  &\simeq
     \SI{2.73e-22}{sec^{-1}} \times
       \sin^2 \theta \times
       \bigg( \frac{m_4}{\si{keV}} \bigg)^5
  \label{eq:nu-nu-gamma-Dirac}
\intertext{in the case of Dirac neutrinos, and \cite{Pal:1981rm, Shrock:1982sc, Xing:2011}}
  \Gamma(\nu_4 \to \nu_a \gamma)^\text{M}
  &= \frac{9 \alpha_\text{em} G_F^2 m_4^5}{256 \pi^4}
       \sum_{j=1,2,3} \bigg( 1 - \frac{m_j^2}{m_4^2} \bigg)^3
       \bigg\{
         \bigg( 1 + \frac{m_j^2}{m_4^2} \bigg)^2
         \bigg[ \sum_{\alpha=e,\mu,\tau}
           \bigg( 1 - \frac{m_\alpha^2}{2 M_W^2} \bigg) \im(U_{\alpha 4} U_{\alpha j}^*) \bigg]^2
                                   \nonumber\\[0.2cm]
  &\qquad
       + \bigg( 1 - \frac{m_j^2}{m_4^2} \bigg)^2
         \bigg[ \sum_{\alpha=e,\mu,\tau}
         \bigg( 1 - \frac{m_\alpha^2}{2 M_W^2} \bigg) \re(U_{\alpha 4} U_{\alpha j}^*) \bigg]^2
       \bigg\}
                                   \nonumber\\[0.2cm]
  &\simeq
     \SI{5.46e-22}{sec^{-1}} \times
       \sin^2 \theta \times
       \bigg( \frac{m_4}{\si{keV}} \bigg)^5
  \label{eq:nu-nu-gamma-Majorana}
\end{align}
for Majorana neutrinos. In these expressions, $m_j$ ($j=1..4$) are the neutrino
mass eigenvalues,  $m_e$, $m_\mu$, and $m_\tau$
denote the charged lepton masses, $M_W$ is the $W$ boson mass, $G_F$ is the
Fermi constant, and $\alpha_\text{em}$ is the electromagnetic fine structure
constant.  The numerical approximations in
\cref{eq:nu-nu-gamma-Dirac,eq:nu-nu-gamma-Majorana} were obtained in the limit
$m_j \ll m_4$ and $m_\alpha \ll M_W$.  Moreover, in this limit, the dependence on
the mixing matrix elements can be expressed in terms of the effective mixing angle
$\sin^2\theta \equiv \sum_j |U_{s4} U_{sj}^*|^2$. Assuming that $\nu_4$ mixes
predominantly with only one of the light mass eigenstates, and that the corresponding
mixing angle is $\ll 1$, $\theta$ can be identified with that mixing angle.
The fact that the expression is different
for the two cases comes from the fact that, for Dirac neutrinos, only an
$\ell^-$ and a $W^+$ can propagate in the loop (opposite for Dirac
antineutrinos), while for Majorana neutrinos, also the combination $\ell^+$ and
$W^-$ is possible.

The radiative decay mode implies that, in spite of its small rate (much smaller
than the inverse age of the Universe), sterile neutrino DM
leads to potentially observable, nearly monoenergetic $\mathcal{O}(\text{keV})$ X-ray
emission in regions of high DM density (Galactic Center, galaxy clusters, etc.).
We call the emission ``nearly monoenergetic'' because the DM velocity dispersion
induces Doppler broadening.  For DM in a galaxy cluster, with a typical velocity
dispersion of order $v \sim \SI{1000}{km/sec}$, the relative line width is
$\sqrt{(1+v)/(1-v)} - 1 \sim 0.3\%$.  Searches for mono-energetic X-rays have been
carried out, and results will be discussed below.

\section{Constraints on Sterile Neutrino Dark Matter}

We have seen in \cref{eq:nu-nu-gamma-Dirac,eq:nu-nu-gamma-Majorana} that the
decay rate of sterile neutrino DM is tiny. However, a modern X-ray
telescope sees about $10^{78}$ dark matter particles in its line of sight to a
nearby galaxy cluster of mass $\SI{e15}{M_{\odot}}$, so a signal may be detectable.

\begin{figure}[t]
  \centering
  \includegraphics[width=0.55\textwidth]{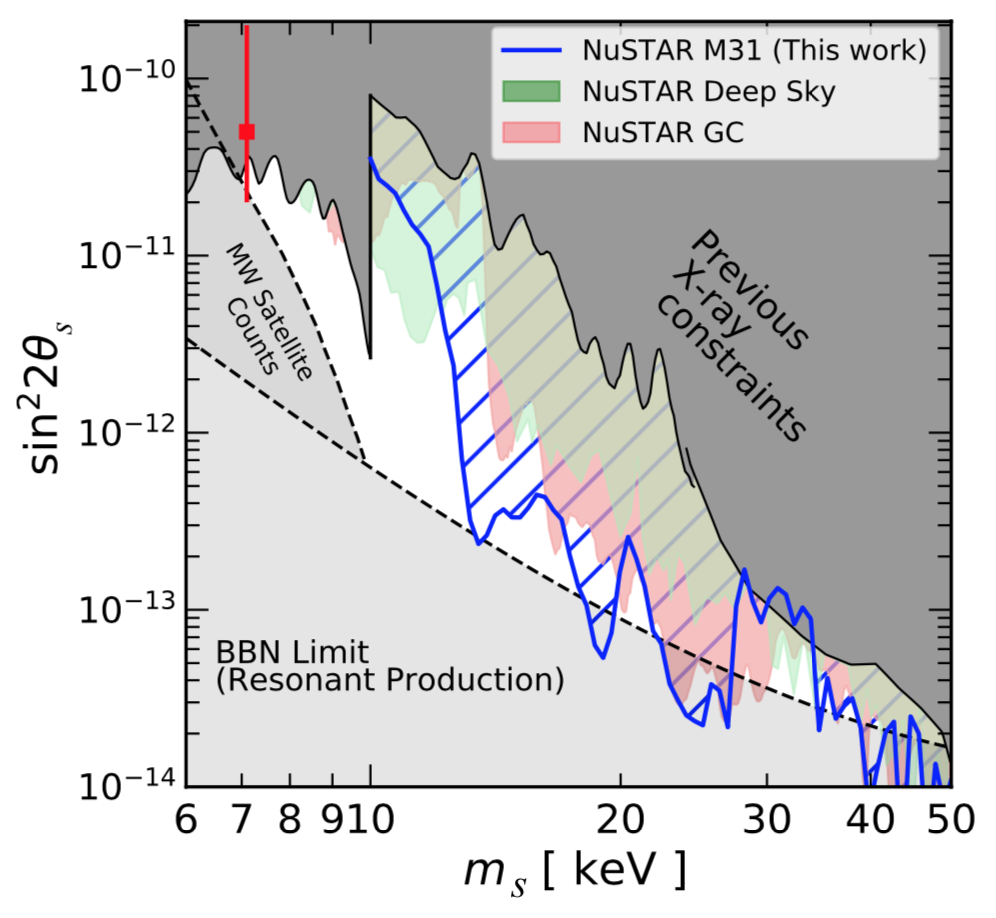}
  \caption{Constraints on sterile neutrino dark matter. Figure taken from~\cite{Ng:2019gch}.
    The colored and gray regions show limits from X-ray searches, while the medium gray
    region on the left labeled ``MW Satellite Counts'' is based on structure formation
    arguments \cite{Cherry:2017dwu}. The region labeled ``BBN Limit (Resonant Production)''
    is disfavored by
    BBN constraints on the lepton asymmetry if the latter is invoked to enhance
    sterile neutrino production. The red dot with an error bar indicates the parameters
    corresponding to the sterile neutrino explanation of the \SI{3.5}{keV}
    line~\cite{Bulbul:2014sua, Boyarsky:2014jta}.}
  \label{fig:sterile-nu-dm-bounds}
\end{figure}

The resulting constraints on sterile neutrino dark matter are summarized in
\cref{fig:sterile-nu-dm-bounds}. We see that only mixing angles as small as
$\sin^2 2\theta \lesssim 10^{-11}$ are still allowed. Even for these, the parameter
space is very limited. It is also important to note that, for such small mixing
angles, the Dodelson--Widrow mechanism can no longer produce the observed
DM abundance. There are alternative mechanisms, though, that can populate this
region of parameter space, including for instance production in the decay of
heavy particles, or production through resonant oscillations. The latter mechanism,
called the Shi--Fuller mechanism \cite{Shi:1998km}, assumes that the lepton
asymetry of the Universe is sizeable (much larger than the baryon asymmetry); in this
case, neutrinos feel an extra Mikheyev--Smirnov--Wolfenstein (MSW) potential
which can enhance the effective mixing
angle in the early Universe, while keeping the vacuum mixing angle relevant
to observations today small.  The problem is that large lepton asymmetries are
difficult to achieve in baryogenesis/leptogenesis models, and they are moreover
constrained by BBN. This constraint is shown in light gray at the bottom of
\cref{fig:sterile-nu-dm-bounds}.  Because smaller mixing angles require
larger lepton asymmetries in the Shi--Fuller mechanism, the BBN bound is
actually a \emph{lower} bound on the mixing angle.

A further constraint arises from large-scale structure formation. Namely,
DM particles with $\mathcal{O}(\text{keV})$ masses are still moving
relatively fast at the time when structure formation starts (matter--radiation
equality, $T \sim \si{eV}$).  They can therefore wash out small-scale density
inhomogeneities, and this affects the subsequent formation of galaxies. In
particular, small structures like dwarf galaxies are then less likely to
be produced. Based on the observed counts of dwarf galaxies accompanying the
Milky Way, one obtains the constraint shown in medium gray in the left part
of \cref{fig:sterile-nu-dm-bounds}. Note that the constraint as shown
here is only valid for sterile neutrinos produced via the Shi--Fuller mechanism.
A similar limit could also be derived for other production mechanisms,
but because each production mechanism leads to a different energy distribution
for the sterile neutrinos at production, the constraint depends on
the production mechanism.

Sterile neutrino dark matter is also constrained by phase space arguments: if
its mass was too low ($\lesssim \SI{1}{keV}$), the conservation of phase space
density would forbid the formation of compact galactic cores.  This constraint
is called the Tremaine--Gunn bound~\cite{Tremaine:1979we}. A slightly weaker
bound arises also from the Pauli exclusion principle, which prohibits
arbitrarily dense packing of fermionic dark matter, again in conflict with
observations of galactic cores. Finally, observations of cosmic structure on
relatively small scales ($\lesssim \SI{10}{Mpc}$) using Lyman-$\alpha$ forests
constrain keV-scale dark matter \cite{Baur:2017stq}, though the quantitative
power of these constraint depends on the dark matter production mechanism.

\section{The 3.5\,keV Anomaly}

\begin{figure}
  \centering
  \includegraphics[width=0.7\textwidth]{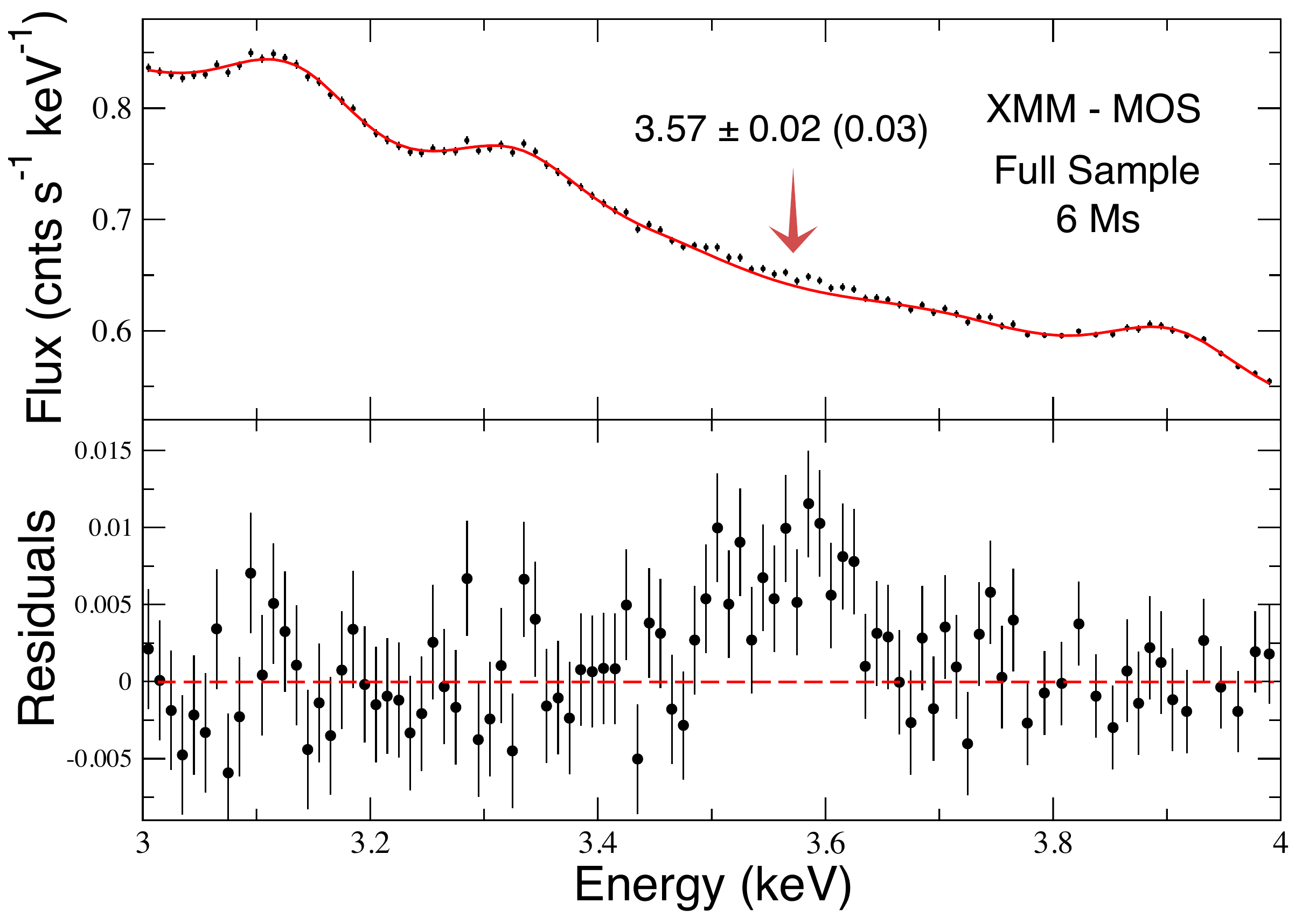}
  \caption{Stacked XMM-Newton spectra from a number of galaxy clusters, leading
  to the detection of an unexplained feature at around \SI{3.5}{keV}. Plot
  taken from \cite{Bulbul:2014sua}.}
  \label{fig:3.5}
\end{figure}

In 2014, stacked observations of galaxy clusters using data from the XMM-Newton
X-ray telescope have led to the detection of an unidentified X-ray line
near 3.55\,keV~\cite{Bulbul:2014sua}, see \cref{fig:3.5}.  (The width of the
excess is compatible with the instrumental resolution, so calling the excess
a line is justified.)

The possibility that this line constitutes a detection of the radiative decay
of a $\sim \SI{7}{keV}$ sterile neutrino has caused a lot of excitement.
Comparing the observed flux from different astrophysical objects (galaxies,
galaxy clusters, \dots), the scaling with the DM abundance in these objects is
not as perfect as one would hope, but also not bad enough to definitely rule
out new physics as an explanation, see for
instance \cite{Dessert:2018qih}.  There is a heated debate going on about the
trustworthiness of both the positive observations and the null results.
Fortunately, future X-ray telescopes with higher energy resolution should be
able to dicriminate between a dark matter origin of the signal and more mundane
explanation in terms of atomic physics effects. This will be possible because
the precise shape of the line is predicted to be different in the two cases.

\section*{Acknowledgements}
It is a great pleasure to thank the organizers of the Les Houches Summer School
2021 on Dark Matter, the students, and the local staff for making this event
a success.

\paragraph{Further reading.}
While we hope that these brief lecture notes offer a compact introduction to
the subject of sterile neutrinos as dark matter candidates, they necessarily
cannot be fully comprehensive. For instance, we did not comment on the large
number of models that exist beyond the minimal Dodelson--Widrow and Shi--Fuller
scenarios. A much more detailed discussion of this and other aspects can be
found in numerous excellent reviews in the literature, see for instance
refs.~\cite{Kusenko:2009up, Drewes:2016upu, Abazajian:2017tcc,
  Boyarsky:2018tvu, Dasgupta:2021ies}.

\paragraph{Funding information.}
The author's work has been partially supported by the European
Research Council (ERC) under the European Union's Horizon 2020 research and
innovation program (grant agreement No.\ 637506, ``$\nu$Directions''). He has
also received funding from the German Research Foundation (DFG) under grant
No.\ KO~4820/4-1.

\setlength{\bibsep}{6pt}
\bibliographystyle{SciPost_bibstyle}
\bibliography{refs}

\nolinenumbers

\end{document}